\documentclass[12pt]{article}
\usepackage{amsmath}
\usepackage{amssymb}
\usepackage{epsfig}
\usepackage{cite}

\newcommand\npb[3]{{\it Nucl.\ Phys.\ }{\bf B #1} (#2) #3}
\newcommand\plb[3]{{\it Phys.\ Lett.\ }{\bf B #1} (#2) #3}
\newcommand\jhep[3]{{\it J. High Energy Phys.\ }{\bf #1} (#2) #3}
\newcommand\prd[3]{{\it Phys.\ Rev.\ }{\bf D #1} (#2) #3}

\newcommand\prep[3]{{\it Phys.\ Rep.\ }{\bf #1} (#2) #3}
\newcommand\prl[3]{{\it Phys.\ Rev.\ Lett.\ }{\bf #1} (#2) #3}

\newcounter{hran}

\def\as{\alpha_{\mbox{\scriptsize s}}}
\def\bas{\bar{\alpha}_s}
\def\ee{e^+e^-}

\def\om{\omega}
\def\Om{\Omega}
\def\lam{\lambda}
\def\sig{\sigma}
\def\Sig{\Sigma}

\def\tcrit{\theta_{\rm crit}}
\def\Eout{E_{\rm out}}
\def\out{\rm out}

 \def\cS{{\cal S}}
 
 \def\cM{{\cal M}}

\begin{document}

\begin{flushright}
  Bicocca--FT--05--41\\
     hep-ph/0601068\\
     December 2005
\end{flushright}

\begin{center}
{\Large\bf Soft gluon emission at large angles}

\vspace{7mm}

{\large Giuseppe Marchesini}

\vspace{2mm}
{\small Dipartimento di Fisica, Universit\`a di Milano-Bicocca and \\
INFN, Sezione di Milano, Italy}
\end{center}

\vspace{8mm}

\centerline{\small\bf Abstract}
\begin{quote}
  I review recent results on soft gluon radiation at large angles.
\end{quote}

\vskip 1cm

\section{Gribov's legacy on soft radiation}

Although the major interest of Vladimir Gribov \cite{Gribov} in QCD
was on non-perturbative aspects (notably colour confinement), his work
on soft photons or soft pions emission in the pre-QCD era
\cite{Gribov-soft} is the basis for understanding general features of
perturbative QCD dynamics.  In principle soft emission is ``trivial'',
it reduces essentially to classical physics. That is, it depends only
on external charges which are treated as classical currents
(asymptotic, large-time states), thus it does not reveal the internal
short distance structure of emitting systems. 

In spite of this ``trivial'' aspects, universality of soft emission
allows one to compute multi-soft gluon emission distributions and
thus to obtain important information on the QCD properties in general.
The key point is that in computing multi-soft gluon amplitudes one
deals with multi-soft scales. The softest gluon does not ``see'' any
internal structure. However the next-to-soft gluon can ``see'' the
softest one. In this way one can recursively reconstruct the detailed
deep structure of the emitting system up to the primary hard emitting
system.
Therefore, while the result is ``trivial'' in QED since only the
primary emitting charges are involved in soft photon emission, in QCD
instead, due to direct gluon interaction, also the successive emitted
gluons are involved and then successively resolved.

In this talk I discuss various results obtained in the course of years
by exploiting soft gluon emission properties obtained in this way.

\section{Multi-soft gluon amplitudes}

I start by recalling how one obtains \cite{BCM} (see also \cite{DKMT})
the amplitude for the emission of $n$ soft gluons $q_1\cdots q_n\,$,
off a colour singlet dipole of primary hard quark antiquark $p_b,p_b$
as emitted for instance in $\ee$ annihilation at high c.m. energy $Q$
(generalization to other dipoles is straightforward).  The general
colour structure is given as sum of Chan-Paton factors
\begin{equation}
\label{eq:cMn}
\cM_n(p_a,p_b;q_1\cdots q_n)=\sum_{\rm perm.}
\{\lam^{b_{i_1}}\cdots \lam^{b_{i_n}}\}_{\beta \bar \beta}\>
M_n(p_aq_{i_1}\cdots q_{i_n}p_b)\,,
\end{equation}
with $\beta,\bar \beta$ and $b_i$ the quark, antiquark and gluon
colour indices (in $SU(N)$ in general). In $M_n$ ({\it colour ordered
  amplitudes}) the soft gluon momenta $q_i$ are ordered as in the
Chan-Paton factor.

In the soft limit $M_n$ can be obtained by recurrence relations. Take
$q_j$ as the softest among the soft gluons.  One uses eikonal emission
(including three-gluon vertex) and colour conservation and obtains
\begin{equation}
\label{eq:rec} 
M_n(\cdots q_{\ell} \,q_{j}\,q_{\ell'}\cdots)\!=\!
g_s\,M_{n\!-\!1}(\cdots q_{\ell}q_{\ell'}\cdots)\cdot 
{\left(\frac{q^{\mu_j}_{\ell}}{(q_{\ell}q_j)}-
\frac{q^{\mu_j}_{\ell'}}{(q_{\ell'}q_j)}\right), 
\quad q_j}\>\>{\rm softest\>\> gluon}\,.
\end{equation}
Here $\mu_j$ is the Lorentz index of softest gluon and $q_{\ell},
q_{\ell'}$ are the momenta ``near in colour'' to $q_j$ in the
permutation considered (they are soft gluons or primary quark or
antiquark). At this stage no information is needed for the deeper
internal structure of $M_{n-1}$.

The factorization structure \eqref{eq:rec} is completely general and
can be iterated.  At each stage $m\!<\!n$ one factorizes the dipole
contribution of the softest gluon in $M_{m}$ and recovers the entire
colour amplitudes $M_{m\!-\!1}$.  The multi-soft gluon amplitude $M_n$
is finally obtained and expressed in terms of successive (energy
ordered) dipole factors for all soft gluons and the Born amplitude
$M_0$ for the primary emitting system.

A direct consequence of the fact that at each stage one recovers the
{\it entire} colour amplitudes is that one does not need to specify the
gauge frame since, at each stage, one is dealing with gauge invariant
amplitudes and dipole emission factors.

By construction $M_n$ is the leading soft part of the amplitude,
proportional to the inverse of all soft gluon energies $M_n\sim
(\om_1\cdots \om_n)^{-1}$. It is valid in any angular configuration
even away from collinear configurations. 
Therefore the factorization form \eqref{eq:rec} is adequate for our
discussion on physics of large angle soft radiation.

Next-to-leading corrections in the soft limit are known
\cite{next-soft}, see also \cite{DMO}. They are expressed in terms of
a factorized $2$-soft gluon emission current. These corrections are
needed in order to reconstruct the argument of the running coupling
$\as$ entering the distribution as the soft gluon transverse momentum
in the center of mass of the emitting dipole.

For the square of the colour ordered amplitude $M_n$ the recurrence
\eqref{eq:rec} gives \cite{BCM}
\begin{equation}
\label{eq:Mn2}
|M_n(p_a q_1\cdots q_n p_b)|^2\!=\!\left(\frac{\as}{2\pi}\right)^n
\!W_{ab}(q_1\cdots q_n)\cdot |M_0|^2\!,\quad 
W_{ab}(q_1\cdots q_n)\!=\!
\frac{(p_ap_b)}{(p_aq_1)\cdots(q_np_b)}\,.
\end{equation}
The distribution $W_{ab}(1\cdots n)$ contains collinear singularities
when any two partons which are near in colour become parallel.
However, this expression is valid in any angular configuration, even
at large angles (hedgehog configurations).  There are three remarkable
properties for this colour distribution.

The first is that in any energy-ordered configuration for the soft
gluons one obtains the same expression for the distribution 
\eqref{eq:Mn2}. This in spite of the fact that the expression of the
colour ordered amplitude $M_n$, obtained from the recurrence relation
\eqref{eq:rec}, depends on the soft gluon energy-ordering.

The second property is the factorization structure
\begin{equation}
\label{eq:Wfactor}
W_{ab}(1\cdots n)=W_{ab}(\ell)\cdot
W_{a\ell}(1\cdots\ell\!-\!1)\,W_{\ell b}(\ell\!+\!1\cdots n)\,,
\end{equation}
with $\ell$ an arbitrary soft gluon. This branching structure allows
the use of parton language.  The colour-ordered distribution
factorizes into the emission of $\ell$ off the primary $ab$-dipole
with the remaining soft gluons emitted by two multi-dipole with
``primary'' partons $a\ell$ and $\ell b$.

The third property refers to the multi-soft-gluon distribution
obtained in the same way \cite{BCM} in the pure Yang-Mills case
(primary dipole partons $p,\bar p$ are hard gluons and the Chan-Paton
factor in \eqref{eq:cMn} is a trace). One finds that the square colour
amplitude obtained in the soft limit coincides with the square of the
exact MHV amplitude obtained by Parke-Taylor \cite{ParkT}.

The full distribution $|\cM_n|^2$ is obtained by squaring
\eqref{eq:cMn}.  To leading order in $N$ it is given by the sum of
permutations of colour ordered distributions \eqref{eq:Mn2} (colour
indices organized in the planar way).  The non-planar contributions
(subleading in $N$) are given by the product of two colour amplitudes
for different permutations.  They are easily computed from
\eqref{eq:rec} for any fixed $n$, however there is not a close
expression valid for arbitrary values of $n$. As a general feature,
non-planar corrections depend in general on the energy ordering and
are less collinear singular than the planar contribution
\eqref{eq:Mn2}, one of the collinear singularities cancels \cite{BCM}.
For these reasons, in studies in which collinear singularities are
relevant, it is justified to neglect non-planar contributions.
For studies of soft radiation at large angle in which collinear
enhancements are not relevant neglecting non-planar contribution could
be less justified.

In the following I schematically report results on physics of large
angle soft emission.  First, using the multi-soft gluon distributions
in the planar approximation, I deduce the generating functional which
provides a ``complete'' description of soft radiation emission.  It
can be formulated as a Markov process and then I describe the
corresponding Monte Carlo simulation procedure.
I describe the use of the generating functional to study the inclusive
distribution of energy deposited in a region away from jets. This
distribution takes contributions, to leading order, from successive
soft gluon branching thus involves the full structure of the multi-soft
gluon distribution.
Finally I discuss distributions for which the planar approximation is
not necessary.  In particular I discuss the fifth form factor in
hadron collision and the non-relativistic heavy quark multiplicity.

\section{Generating functional (soft and planar limit)}
Complete information on the emission of soft gluons $\{q_1\ldots
q_n\}$ off the dipole $p_ap_b$ is contained in the generating
functional
\begin{equation}
\label{eq:Sig}
\Sigma_{ab}[Q,u]=\sum_n\!\frac{1}{n!}
\int\!\frac{d\sigma_{ab}^{(n)}}{\sigma_{ab\,T}}
\prod_{i=1}^n\! u(q_i)\,,
\end{equation}
with $Q$ the hard scale, $\sigma_{ab}$ the dipole distributions and
$u(q)$ the source which serves to specify the radiation observable
considered. In the planar approximation the multi-soft gluon
distribution $d\sig^{(n)}_{ab}(q_1\cdots q_n)$ is proportional to
$W_{ab}(q_1\cdots q_n)$.  Taking the derivative with respect to $Q$
(with energy ordered phase space) and using the branching
factorization property \eqref{eq:Wfactor} one obtains the evolution
equation \cite{BMS}
\begin{equation}
\label{eq:eveq}
Q\partial_{Q}\Sigma_{ab}[Q,u]\!=\!\int\! \frac{d\Om_q}{4\pi}\, 
\bas w_{ab}(q) \big\{
{u(q)}\,\Sigma_{aq}[Q,u]\!\cdot\!\Sigma_{qb}[Q,u]\!-\!\Sigma_{ab}[Q,u]
\big\}, 
\end{equation}
with 
\begin{equation}
   \label{eq:wab}
 \bas=\frac{N\,\as}{\pi}\,,\qquad
   w_{ab}(q)=\om_q^2\!\cdot\!W_{ab}(q)=\frac{\xi_{ab}}{\xi_{aq}\xi_{qb}}\,,\qquad
\xi_{ij}\!=\!1\!-\!\cos\theta_{ij}\,.
\end{equation}
The initial condition is set at a scale $Q_0$ corresponding to no
emission, $\Sigma_{ab}[Q_0,u]=1$.  Virtual corrections (the last term
in \eqref{eq:eveq}) are included to satisfy unitarity
$\Sigma[Q,u\!=\!1]\!=\!1$. This realizes the real-virtual cancellation
of the collinear singularity of $w_{ab}(q)$ for $\xi_{aq}\!\to\!0$ or
$\xi_{qb}\!\to\!0$. This evolution equation reflects the factorization
branching structure \eqref{eq:Wfactor} of the multi-dipole
distribution.

The generating functional \eqref{eq:Sig} fully describes soft emission
even at large angles and thus it is valid beyond the collinear
singularity approximation.  However is difficult to generalize the
evolution equation \eqref{eq:eveq} beyond soft approximation including
non-soft emission and recoil.

In the next Sections I will report applications exploiting
\eqref{eq:eveq}.

\section{Large angle soft emission and Monte Carlo} 

The evolution equation \eqref{eq:eveq} for $\Sig_{ab}$ can be solved
numerically by Monte Carlo simulation. In the simulation one generates
events with any number of soft gluons which can be used to compute any
final state observables.  First one needs to resum virtual
corrections. To split the real and virtual terms in \eqref{eq:eveq}
one introduces a cutoff $Q_0$ in the dipole transverse momentum.  The
resummation of virtual corrections gives rise to the Sudakov form
factor
\begin{equation}
  \label{eq:sud}
\begin{split} 
\ln S_{ab}(Q,Q_0)=-\!\int_{Q_0}^Q\!
\frac{d\om_q}{\om_q}\,\frac{d\Om_q}{4\pi}\,\bas(q_{ab\,t})\,w_{ab}(q)\,
\cdot\theta(q_{ab\,t}\!-\!Q_0),\qquad
q_{ab\,t}^2=\frac{2\om_q^2}{w_{ab}(q)}\,.
\end{split}
\end{equation}
The evolution equation \eqref{eq:eveq} can be written as
\begin{equation}
  \label{eq:branch}
\Sig_{ab}[Q,u]=S_{ab}(Q,Q_0)+
\int_{Q_0}^Q\,dP_{ab}(q)\,u(q)\,\Sig_{aq}[\om_q,u]\cdot\Sig_{qb}[\om_q,u])\,,
\end{equation}
with $dP_{ab}(q)$ the splitting probability which can be casted in
the following factorized form
\begin{equation}
  \label{eq:dP}
dP_{ab}(q)=d\left(\frac{S_{ab}(Q,Q_0)}{S_{ab}(\om_q,Q_0)}\right)
\cdot dR_{ab}(\Om_q)\,,\qquad \om_q<Q\,,
\end{equation}
with
\begin{equation}
  \label{eq:dR}
\frac{dR_{ab}(\Om_q)}{d\Om_q}=\frac{
\bas(q_{ab\,t})\,w_{ab}(q)}{N_{ab}(\om_q)}
\cdot\theta(q_{ab\,t}\!-\!Q_0)\,,\qquad
\int dR_{ab}(\Om)=1\,.
\end{equation}
From \eqref{eq:branch} one has that the Sudakov form factor
$S_{ab}(Q,Q_0)$ can be interpreted as probability for no-emission of a
soft gluon with dipole transverse momentum larger than $Q_0$.

Using this branching structure and the factorized form of the
splitting probability \eqref{eq:dP} one can construct a Monte Carlo
event generator (see also \cite{LL}).  To generate an event (with
resolution $Q_0$) one proceeds as follows.  Starting from the
$ab$-dipole with virtuality $Q$, one tries to emit a soft gluon $q$
using \eqref{eq:dP}. To this end one first sorts a random number
$0\!<\!r\!<\!1$:
\begin{itemize} 
\item for $r\!<\!S(Q,Q_0)$ the dipole does not branch (with $Q_0$ cutoff);
\item for $r\!>\!S(Q,Q_0)$ a soft gluon is emitted with energy $\om_q$ 
obtained by solving 
\begin{equation}
{S_{ab}(\om_q,Q_0)}\cdot r={S_{ab}(Q,Q_0)}\,.
\end{equation}
\end{itemize}
Then, using \eqref{eq:dR}, one finds the direction $\Om_q$.  
If the emission takes place, one has to deal with the two dipoles $aq$
and $qb$ which are further emitting soft gluons with energy smaller
than $\om_q$.  The procedure is repeated for each dipole till no
dipole is emitting (withing the cutoff $Q_0$).
Such an event generator can be used to study radiation quantities in
the leading soft limit. Due to strong ordering, no recoil is
considered.  
To make a realistic simulation valid beyond soft limit one needs to
include parton recoil, full collinear structure and quark branching.
Finally hadronization had to be accounted for.

A way to account for non-soft emission consists in taking the
collinear limit of \eqref{eq:Mn2} and one obtains \cite{BCM} the
evolution equation in angles (instead of in energy). Using angular
evolution one can easily introduce recoil, the full splitting
functions (including quarks and the full $N$ dependence) and initial
state emission. In the angular evolution large angle soft emission are
accounted for only in an average sense (coherence through angular
ordering). The angular evolution is the basis for the Monte Carlo
simulations \cite{MC} presently used.  There is a continuous upgrading
of these Monte Carlo simulations \cite{MCupgr}.  It would be important
to match the angular evolution with the one which fully accounts for
large angle soft emission as given in \eqref{eq:eveq}.

\section{Large angle soft emission and jet characteristics} 

Measuring final state characteristics in hard interactions supplements
the overall hardness scale $Q$ with the second scale $Q_0\ll Q$ that
quantifies small deviation of the final state system from the Born
kinematics.  The ratio of these two scales being a large parameter
calls for analysis and resummation of double (DL) and single
logarithmic (SL) radiative corrections in all orders.  Logarithmically
enhanced (both DL and SL) contributions of {\em collinear}\/ origin
take contributions from bremsstrahlung radiation off primary hard
partons and are resummed into exponential Sudakov form factors.  SL
effects due to soft gluons radiated at large angles, when present,
pose more problems. In the following I discuss two cases in which
these SL contributions are actually present.

The first case refers to the so called {\it non-global observables}
that acquire contributions from a restricted phase space window.
Examples are particle energy flow $E\!=\!Q_0$ in a given inter-jet
direction, Sterman-Weinberg distribution (energy in a cone), photon
isolation, rapidity cuts in DIS, hadron-hadron inter jet string/drag
effects, profile of a separate jet (current hemisphere).  

The second case consists in measuring final state characteristics in
hard hadron--hadron interactions.

\subsection{Non-global logarithms}

Non-global observables acquire SL contributions from ensembles of
$n$-energy ordered gluons radiated at arbitrary (large) angles which
are resummed \cite{DS}, at the planar level, by the soft gluon
evolution equation \eqref{eq:eveq}.  The simplest case is the study
in $\ee$ of the radiation emitted in the region ``out'' away from the
thrust axis as shown here
\begin{center}
   \epsfig{file=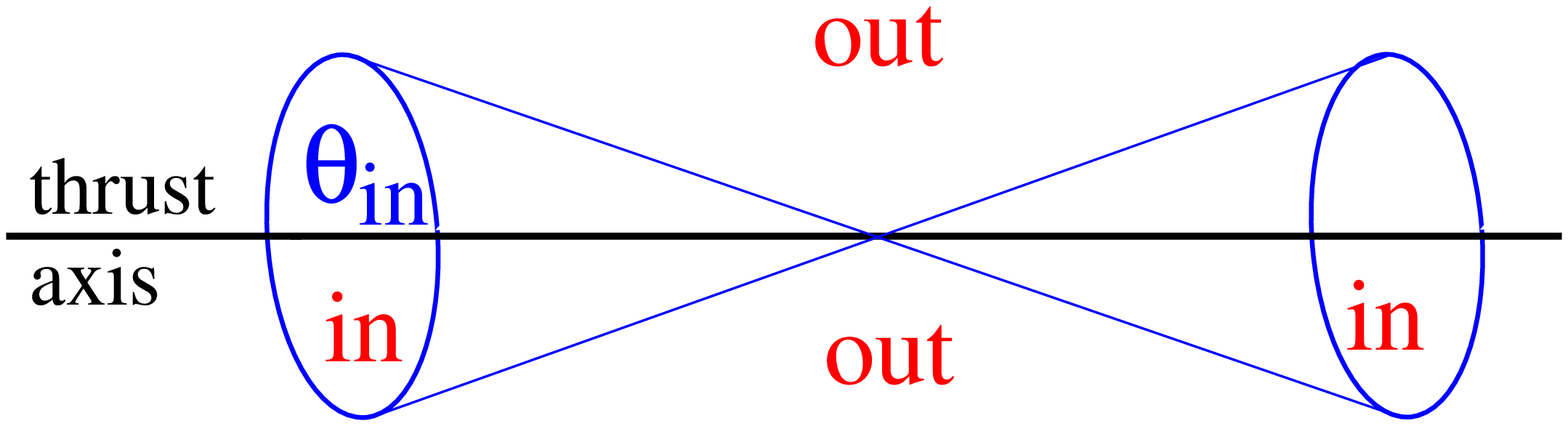,width=.5\textwidth}
\end{center}

\noindent
Most of emission is contributing to the two jets around the thrust axis.
Therefore, to study the distribution in energy $\Eout$ away from jet
\begin{equation}
\Sigma_{\rm out}(Q,\Eout)\!=\!\sum_n\!\int\frac{d\sigma_n}{\sigma_T}\>
\Theta\!\left(\!\Eout\!-\!\sum_{ \out}q_{ti}  \right)
\end{equation}
one needs a careful description of soft gluons emitted at large angle,
that is the transverse momentum evolution equation \eqref{eq:eveq}.
Generalizing to a the emission off a general $ab$-dipole and
specializing the soft gluon source $u(q)$ to this specific observable,
one obtains \cite{BMS}
\begin{equation}
\label{eq:eveq-Eout}
\partial_{\tau}\Sigma_{ab}(\tau)=
-(\partial_{\tau}R_{ab})\!\cdot\!\Sigma_{ab}(\tau)\>+
\int_{\rm { in}}\!\frac{d\Omega_q}{4\pi}\, w_{ab}(q)
\left[\Sigma_{aq}(\tau)\!\cdot\!\Sigma_{qb}(\tau)\!-\!\Sigma_{ab}(\tau)\right],
\end{equation}
where, at SL accuracy, the distribution $\Sig_{ab}(\tau)$ is a
function of the two scales $Q$ and $\Eout$ via the logarithmic
integral of the running coupling
\begin{equation}
  \label{eq:tau}
  \tau=\int_{Q_0}^Q\frac{dq_t}{q_t}\bas(q_t)\,,\qquad Q_0=\Eout\,,
\end{equation}
and $R_{ab}(\tau)$ the $ab$-radiator evaluated in the region away from
the jets (region ``out'')
\begin{equation}
  \label{eq:Rab}
R_{ab}(\tau)=\int_{\Eout}^Q\frac{dq_t}{q_t}\bas(q_t)
\int_{\rm{out}}\!\frac{d\Omega_q}{4\pi}\,w_{ab}(q)
\>\simeq\> 2\tau\,\ln\frac{2}{\theta_{\rm in}}\,.
\end{equation}
In \eqref{eq:eveq-Eout} the radiator term corresponds to
bremsstrahlung emission while the integral term corresponds to the
soft gluon branching which takes place inside the jet region (region
``in'') and gives rise to ``non-global logs''.  The solution of
\eqref{eq:eveq-Eout} has the form
\begin{equation}
\label{eq:out}
\Sigma_{ab}(\tau)=e^{-R_{ab}(\tau)}\cdot C_{ab}(\tau).
\end{equation}
The first is the Sudakov form factor.  The second factor $C_{ab}$
comes from the branching inside the jet region\footnote{The integral
  term in \eqref{eq:eveq-Eout}, and then the correlation $C_{ab}$, is
  absent for global observables which involve the full phase space.}.
Collinear singularities in $w_{ab}(q)$ partially cancel between real
and virtual contributions.  At large $\tau$ virtual corrections are
overwhelming real contributions and one finds the Gaussian behaviour
\begin{equation}
\label{eq:Cab}
C_{ab}(\tau)\sim e^{-\frac{c}{2}\,\tau^2}\,,\qquad c=4.883...
\end{equation}
The asymptotic beheviour \eqref{eq:Cab} is independent on the geometry
of the recorded region away from the jet.  The value of the constant
$c$ is related to universal features of the evolution equation in the
asymptotic regime and enters also in the small $x$ physics as we shall
see. In \eqref{eq:out} the Sudakov form factor is negligible with
respect to the correlation function at large $\tau$.

Berger, K\'ucs and Sterman \cite{BKS} have formulated a programme of
how to avoid non-global logarithms in a measurement of transverse
energy flow away from jets. They suggested to squeeze the jets by
taking the thrust $T$ close to one and thus suppress multi-gluon
branching inside the jet region. They introduced the {\it jet/shape
  correlation} $\Sigma _{fs}(Q,V,\Eout)$ in the two variables, the
energy-flow $\Eout$ and the global jet-shape $V$, for instance
$V\!=\!1\!-\!T$.  They showed that for $V\sim \Eout/Q\ll1$ the
flow/shape correlation reduces to the global jet-shape distribution
$\Sigma(Q,V)$.  This result has been extended \cite{DM-out} for general
(small) values $V$ and $E_{\rm out}/Q$. At SL accuracy one finds the
factorized result
\begin{equation}
\Sigma _{fs}(Q,V,\Eout)\simeq\Sigma(Q,V)\cdot\Sigma_{\out}(VQ,\Eout) \,,
\end{equation}
so that the associated measure is equivalent to the two independent
ones, the global one $\Sigma(Q,V)$ at the scale at $Q$ and the
non-global one at $VQ$.

Analysis of non-global observables can be generalized (in planar
approximation) to hard hadron-hadron collisions. In planar
approximation, the hard square matrix element could be expressed as a
superposition of dipoles and each $ab$-dipole contributes with the
distribution $\Sig_{ab}(\tau)$ in \eqref{eq:out}.  In hadron
collision, contributions beyond planar approximations have been
obtained only for global jet-shape observables as we shall discuss in
the following.

\subsection{Hadron interaction and the fifth form factor}

We discuss now global jet characteristics in hard hadron--hadron
interaction with the underlying parton scattering process $p_1,p_2\to
p_3 ,p_4$ at a scale $Q$. Examples are out-of-event-plane energy
production, near-to-backward particle correlations, inter-jet energy
flows, etc. They are characterized by a scale $Q_0\!\ll\! Q$.

For global observables all enhanced contributions at DL and SL level
comes from bremsstrahlung radiation off the four primary partons.
This implies that the problem reduces to the analysis of {\em virtual
  corrections}\/ due to multiple gluons with $q_t>Q_0$ attached to
primary hard partons only.  They can be treated iteratively and fully
exponentiated.

In addition to the standard collinear contributions leading to Sudakov
for factors associated to the four primary partons $p_i$, one has SL
enhancements coming from large angle soft emission. This is due to the
fact that soft gluon emission changes the {\em colour state}\/ of the
hard parton system which in turn affects successive radiation of a
softer gluon.  Resummation of these soft logarithms generates a fifth
form factor in addition to the Sudakov form factors.

The programme of resumming soft SL effects due to large angle gluon
emission in hadron--hadron collisions was pioneered by Botts and
Sterman~\cite{BS} and discussed in general also in \cite{BCMN}.  The
treatment for finite $N$ of large angle gluon radiation in global
final state characteristics is based on the observation that the
square of the eikonal current for emission of a soft gluon $q$ off the
ensemble of four hard partons $p_i$ can be represented as
\begin{equation}
\label{eq:alter}
\begin{split}
  - j^2(q) =\> & T_1^2\, W^{(1)}_{34}(q) + T_2^2\, W^{(2)}_{34}(q) +
  T_3^2\, W^{(3)}_{12}(q) +  T_4^2\, W^{(4)}_{12}(q)  \\
  &+\> T_t^2\cdot A_{t}(q) \> \>+ T_u^2\cdot A_u(q)\,.
\end{split}
\end{equation} 
Here $T_i^2$ is the $SU(N)$ ``colour charge'' of parton $p_i$ and the
two operators $T_t^2\!=\!(T_1\!-\!T_3)^2$ and
$T_u^2\!=\!(T_1\!-\!T_4)^2$ are the charges exchanged in the $t$- and
$u$- channels.  In the first line one has the combination
$W^{(c)}_{ab}\! =\!  w_{ac}+w_{bc}-w_{ab}$ with the singular integral
\begin{equation}\label{eq:W123def}
  \int\frac{d\Omega}{4\pi}\> W ^{(c)}_{ab} \>=\> 
\ln\frac{Q^2}{2m^2}\,,\quad Q^2=\frac{tu}{s}=s\sin^2\Theta_s\,,
\end{equation}
with $s\!=\!(p_1\!+\!p_2)^2$, $t\!=\!(p_1\!-\!p_3)^2$ and
$u\!=\!(p_1\!-\!p_4)^2$.  The distribution $W^{(c)}_{ab}$ is collinear
singular {\em only}\/ when $q$ is parallel to $p_c$ Here $m^2$ is a
collinear cutoff which is replaced by the proper observable dependent
scale ${Q_0}$ when real and virtual contributions are taken together.
Its exponentiation leads to the Sudakov form factors $F_c(Q_0,Q)$ with
colour charge of parton $c$ and common hard scale $Q$ in
\eqref{eq:W123def}.

The last two terms in \eqref{eq:alter} are collinear regular 
\begin{equation}
\label{eq:At}
A_t\!=\! w_{12}\! +\!w_{34}\!-\!w_{13}\!-\!w_{24}\,,\qquad 
\int\frac{d\Omega}{4\pi}\, A_t(q) \>=\>  2\ln\frac{s}{-t}\,,
\end{equation} 
and similarly for $A_u$. Non-Abelian Coulomb corrections due to
virtual gluons exchanges between the two incoming or two outgoing
partons can be simply incorporated by replacing $-t,-u$ with $t,u$ in
\eqref{eq:At}.  Contrary to the first four DL contributions, this
additional contribution originates from coherent gluon radiation at
angles {\em larger than the cms scattering angle $\Theta_s$}. It gives
rise to the {\em fifth form factor} which is a single logarithmic
function of $\tau$ given in \eqref{eq:tau} with $\Eout$ replaced by
$Q_0$.

In summary, the two scale distributions ais given by
\begin{equation}
   \Sigma(Q_0,Q) \>=\> \Sigma^{{\mbox{\scriptsize coll}}}(Q_0,Q) \cdot
   \cS_X(\tau), 
\end{equation}
Here $\Sigma^{{\mbox{\scriptsize coll}}}$ embodies the first four
terms in \eqref{eq:alter} as well as collinear logarithms from parton
distribution functions.  $\cS_X(\tau)$ is the {\em soft factor} coming
from the fifth form factor.

In the following I illustrate the solution in some relevant special
cases for gluon--gluon scattering in $SU(N)$ that was first treated
in~\cite{KOS}.  The answer for the soft factor $\cS_X$ is expressed in
terms of the following suppression factors
\begin{equation}
  \chi_t(\tau) \>=\> \exp\left\{-2N\tau \cdot\ln\frac{s}{-t}\right\},
  \quad \chi_u(\tau) \>=\> \exp\left\{-2N\tau
  \cdot\ln\frac{s}{-u}\right\}.
\end{equation}

\paragraph{Scattering at $90^{\rm o}$.}
In this kinematical configuration ($t\!=\!u\!=\!-\!s/2$) one has
\begin{equation} 
\label{eq:Sb0}
 \cS_X(\tau) = \frac{\chi^2}{3}\left[\, \frac{4\, \chi^{2}}{N^2\!-\!1} +
 { \chi} + \frac{N\!-\!3}{N\!-\!1}\, \chi^{\frac2N} +
 \frac{N\!+\!3}{N\!+\!1}\,
\chi^{-\frac2N} \,\right], \quad \chi(\tau)\,=\,
\exp\left\{-2N\tau\ln2\right\}.
\end{equation}

\paragraph{$N\to\infty$ limit.}
The soft factor becomes
\begin{equation}
\label{eq:x0ans}
  \cS_X(\tau) \>=\> {\chi_t \, \chi_u}\> \frac{ (m_t+m_u)^2 +
  (m_s-m_u)^2\,\chi_t + (m_s+m_t)^2\,\chi_u}{(m_t+m_u)^2 + (m_s-m_u)^2
  + (m_s+m_t)^2}.
\end{equation}
Here $m_s$, $m_t$, $m_u$ are pieces of the Born $gg$ scattering matrix
element each containing the gluon exchange diagram in the
corresponding channel (together with the piece of the four-gluon
vertex with the same colour structure). One has
$$
 (m_t+m_u)^2 \>=\>  1-\frac{st}{u^2}-\frac{us}{t^2} +\frac{s^2}{tu} ,
$$
with the other two obtained by crossing symmetry. 

\paragraph{Regge limit}
In the case of close to forward scattering, $|t|\ll s\simeq |u|$.
Since for $t\to 0$ scattering in the Born approximation is dominated
by $t$-channel one gluon exchange, we are left with 
\begin{equation}
\label{eq:b1ans}
    S(\tau)\>=\> \chi_t(\tau) \>=\> \left(\frac{s}{t}\right)^{-2N\tau},
\end{equation}
which exponent coincides with the (twice) Regge trajectory of the
gluon.

The eigenvalues of the anomalous dimension matrix possesses a weird
symmetry which interchanges of internal (colour group) and external
(scattering angle) degrees of freedom:
\begin{equation} \label{eq:weird2}
  \frac{\ln s^2/tu}{\ln t/u} \quad \Longleftrightarrow \quad  N\,.
\end{equation} 
In particular, this symmetry relates $90$-degree scattering, $t=u$,
with the large-$N$ limit of the theory.  Giving the complexity of the
expressions involved, such a symmetry being accidental looks highly
improbable. Its origin remains mysterious and may points at existence
of an enveloping theoretical context that correlates internal and
external variables (string theory?).

\section{Is there a BFKL Pomeron in jet emission?}
It is surprising that large angle soft emission in jet physics has
some relation with high-energy scattering and BFKL dynamics. This in
spite of the difference in the relevant multi-soft gluon kinematical
configurations.  In the following I recall two examples in which the
two dynamics are (formally) related.

\paragraph{Kovchegov equation.} 
The evolution equation \eqref{eq:eveq-Eout} for the jet observable
$\Sigma_{ab}(\tau)$ has some resemblance with the Kovchegov equation.
To make the resemblance more striking, take a $ab$-dipole with small
angle $\theta_{ab}\!\ll\!1$ and introduce the two-dimensional
``angular'' variable $\vec{\theta}_{ab}$ and $\Sigma_{ab}(\tau)
\!=\!\Sigma(\tau,\vec{\theta}_{ab}) $.  The evolution equation
\eqref{eq:eveq} becomes
\begin{equation}
\label{eq:eveqS}
\partial_{\tau}\Sigma(\tau,\vec{\theta})=
\int\frac{d^2\theta'}{2\pi}
\frac{{\theta'}^2}{{\theta'}^2(\vec{\theta}\!-\!\vec{\theta}')^2}
\Big[\Sigma(\tau,\vec{\theta'})\Sigma(\tau,\vec{\theta'}\!-\!\vec{\theta})
-\Sigma(\tau,\vec{\theta})\Big].
\end{equation}
Here also the angular integration has been replaced by the
two-dimensional integration over $\vec{\theta}'$. One recognizes that
\eqref{eq:eveqS} formally coincides with the Kovchegov equation for
elastic $S$-matrix at high energy, $S(\tau,\vec{x})$ where $\vec{x}$
is the impact parameter and $\tau =N\,\as/\pi\,Y$ with $Y$ the
rapidity and $\bas$ the fixed coupling.  For large $\tau$ one has (see
the asymptotic behaviour of the correlation in \eqref{eq:Cab})
\begin{equation}
\label{eq:large-tau}
\Sigma(\tau,\vec{\theta})\sim e^{-\frac{c}{2}\tau^2}\,, \quad\mbox{for}\>\> 
\theta > \tcrit(\tau)\sim e^{-\frac{c}{2}\tau}\,,\quad c=4.883...
\end{equation}
so that virtual corrections are overwhelming above $\tcrit$ (buffer
region for the jet distribution \cite{DS}).  The asymptotic behaviour
\eqref{eq:large-tau} corresponds, in the case of the Kovchegov
equation, to the saturation for the $S$-matrix with $1/\tcrit(\tau)\to
Q_s(\tau)$, the saturation scale.  The parameter $c$ is determined
\cite{MT} by the characteristic function of the BFKL equation which is
the linear limit of \eqref{eq:eveqS}, as I shall discuss next.

\paragraph{BFKL equation.} 
An other case of a formal connection between high energy and jet
physics is encountered \cite{MM} considering the multiplicity of
  non-relativistic heavy $Q\bar Q$ pair of total mass $\cM$ emitted by
  a $ab$-dipole.  This observable $N(\rho_{ab},\tau)$, with
  $\rho_{ab}\!=\!(1\!-\!\cos\theta_{ab})/2$ and $\tau$ given in
  \eqref{eq:tau} for $Q_0$ replaced by $\cM$, satisfies the linearized
  version of \eqref{eq:eveq} which, after azimuthal integration, is
  given by
\begin{equation}
  \label{eq:eveqMM}
  \partial_{\tau} N(\rho,\tau)\!=\!\int_0^1\frac{d\eta}{1-\eta}
  \left(\eta^{-1} N(\eta\rho,\tau)-N(\rho,\tau)\right)
  +\int_{\rho}^1\frac{d\eta}{1-\eta}
  \left(N(\eta^{-1}\rho,\tau)-N(\rho,\tau)\right)\,.
\end{equation}
The only formal difference between \eqref{eq:eveqMM} and the BFKL
equation for the $T$-matrix at high energy is the presence of the
lower limit $\eta\!>\!\rho$ in the second integral which ensures that
the argument of $N(\rho/\eta,\tau)$ remains within the physical region
$\rho/\eta<1$. We have\footnote{The well known result for the
  multiplicity in the collinear limit \cite{BCM,DKMT} is obtained by taking the
  collinear piece in \eqref{eq:eveqMM} giving (for fixed $\as$)
$$
\partial_\tau N_{\rm coll}(\rho,\tau)=\int_0^\rho \frac{d \rho'}{\rho'} \,
N_{\rm coll}(\rho',\tau)\,,\qquad N_{\rm coll}(Q)=
N_0 \exp \sqrt{2\bas}\,\ln Q/Q_0\,,
$$ 
with $Q^2=E^2\rho$ and $Q_0$ a collinear cutoff.}
\begin{equation}
  \label{eq:N-asympt}
N(\rho,\tau)\>\sim\>e^{\tau\cdot 4\ln2}\cdot 
\frac{e^{-\ln^2\rho\,/{2D\tau}}}{\tau\sqrt{2\pi D\tau \ln \rho}}\,.
\end{equation}
Here $4\ln2$ and $D=28\zeta(3)$ are related to the BFKL characteristic
function. The difference of this asymptotic behaviour with the one for
the $T$-matrix in BFKL equation is given by $1/\tau$ in the prefactor
which is the result of the fact that $\rho$ is an angular variable
with constraint $\rho<1$.  For details on the jet physics solution and
explanation on the origin of the additional $1/\tau$ suppression
pre-asymptotic factor in $N(\rho,\tau)$, see \cite{MM}.

The basis for the relation between the equations in jet-physics
(\eqref{eq:eveqS}, \eqref{eq:eveqMM}) and the Kovchegov and BFKL ones
in high-energy scattering is of course the (same) multi-soft
gluon-distribution. However the dominant contributions in the jet and
high energy case come from very different kinematical configurations
as we shall discuss now.

{\it Jet-physics case}. Here all angles of emitted gluons are of same
order. This is due to the fact that the observables considered in this
Section do not contain collinear singularities.  Moreover, in the
(leading) infrared limit soft gluon energies are ordered so that also
the emitted transverse momenta are ordered. The ordered variable
$q_{t}$ enters the argument of the running coupling so that $\tau$ is
never too large.
  
{\it High-energy scattering case}. Here all intermediate soft gluon
transverse momenta are of same order (no singularities for vanishing
transverse momentum differences). On the other hand, energy ordering
implies in this case that intermediate gluon angles are ordered.
Contrary to the previous case, the running coupling depends on
transverse momenta which are all of same order.  Therefore, in first
approximation, one can take $\as$ fixed and so that $\tau$ becomes
large at high energy.

\section{Concluding remarks}

I recalled recent results in the studies of soft gluon emission at
large angles (hedgehog configurations). Here I present some of the
possible developments of these studies.

Previous studies on non-perturbative corrections (power corrections)
to perturbative results has been analysed for quantities involving
emission in the collinear regions. It would be important to study
power corrections for observables involving hedgehog soft gluon
configurations (without collinear enhancements) in which takes place
neutralization of colour between jets takes place.

Present Monte Carlo simulations involves contributions of large angle
soft emission only in an approximate sense (angular ordering). It
would be important to include exact contributions from hedgehog
configuration. This requires to merge the Monte Carlo simulation here
described with the one presently used.

Multi-soft gluon distributions $|\cM_n|^2$ are known for arbitrary $n$
only in the planar approximation. It would be important to have a
close expression of the distribution for any $n$ and arbitrary $N$ or
at least the subleading $1/N$ corrections.  They are specially
important for a correct description of hedgehog configurations.

Similarities and differences in the dynamics of high energy scattering
and jet-physics (with non-global logs) should be further exploited in
order to have new insights in both fields.

\end{document}